\documentclass[]{ceurart}
\usepackage{listings}
\usepackage{longtable}
\usepackage{lscape}
\usepackage{geometry}
\usepackage{float}
\usepackage[justification=centering]{caption}
\usepackage[T1]{fontenc} 

\usepackage{ifthen}

\usepackage{arydshln} 
\usepackage[many]{tcolorbox} 
\usepackage{enumerate}
\usepackage{enumitem}
\usepackage[normalem]{ulem} 			
\usepackage{amsmath,amsfonts}

\usepackage{caption} 
\usepackage{subcaption} 

\usepackage{color}
\usepackage{listings}
\usepackage{etoolbox}
\usepackage{textcomp}

\definecolor{stComment}{rgb}{0.5,0.5,0.5}
\definecolor{stString}{rgb}{0.58,0,0.82}
\definecolor{stKeywords}{rgb}{0.21,0.55,0.7}
\definecolor{stNumbers}{rgb}{.5,0,0}

\newtoggle{InString}{}
\togglefalse{InString}
\newcommand*{\ColorIfNotInString}[1]{\iftoggle{InString}{#1}{\color{stNumbers}#1}}%

\lstdefinelanguage{Pharo}{
  keywordstyle=\color{stKeywords},
  commentstyle=\color{stComment},
  stringstyle=\color{stString},
  alsoletter=\#,
  identifierstyle=\idstyle, 
  showstringspaces=false,
  morekeywords={true,false,self,super,nil},
  sensitive=true, 
  morecomment=[s]{"}{"}, 
  morestring=[d]', 
  style=PharoStyle,
  tabsize=2,
  basicstyle=\small\ttfamily,
  upquote=true,
}

\makeatletter%
\newcommand*\idstyle[1]{%
  \expandafter\id@style\the\lst@token{#1}\relax%
}
\def\id@style#1#2\relax{%
  \ifnum\pdfstrcmp{#1}{\#}=0%
  \ttfamily\color{stString} \the\lst@token%
  \else%
  \edef\tempa{\uccode`#1}%
  \edef\tempb{`#1}%
  \ifnum\tempa=\tempb%
  \ttfamily\color{blue} \the\lst@token%
  \else%
  \the\lst@token%
  \fi%
  \fi%
}

\lstdefinestyle{PharoStyle}{ 
  literate=%
  {^}{{$\uparrow$}}1%
  {0}{{{\ColorIfNotInString{0}}}}1%
  {1}{{{\ColorIfNotInString{1}}}}1%
  {2}{{{\ColorIfNotInString{2}}}}1%
  {3}{{{\ColorIfNotInString{3}}}}1%
  {4}{{{\ColorIfNotInString{4}}}}1%
  {5}{{{\ColorIfNotInString{5}}}}1%
  {6}{{{\ColorIfNotInString{6}}}}1%
  {7}{{{\ColorIfNotInString{7}}}}1%
  {8}{{{\ColorIfNotInString{8}}}}1%
  {9}{{{\ColorIfNotInString{9}}}}1%
} 

\newcolumntype{L}[1]{>{\raggedright\let\newline\\\arraybackslash\hspace{0pt}}p{#1}}
\newcolumntype{C}[1]{>{\centering\let\newline\\\arraybackslash\hspace{0pt}}p{#1}}
\newcolumntype{R}[1]{>{\raggedleft\let\newline\\\arraybackslash\hspace{0pt}}p{#1}}

\newboolean{showcomments}
\setboolean{showcomments}{true}
\usepackage{tikz}
\usepackage{collcell}
\usepackage{tabularx}

\usepackage{quoting,xparse}

\ifthenelse{\boolean{showcomments}}
{
	\newcommand{\nb}[3]{
		{\colorbox{#2}{\bfseries\sffamily\scriptsize\textcolor{white}{#1}}}
		{\textcolor{#2}{\sf\small$\blacktriangleright$\textit{#3}$\blacktriangleleft$}}}
	 
	\newcommand{\bnote}[2]{\fbox{\color{blue}\bfseries\sffamily\scriptsize#1}
    	{\color{blue}\sf\small$\blacktriangleright$\textit{#2}$\blacktriangleleft$}}
	\newcommand{\old}[1]{{\color{gray}\sout{#1}}} 
	\newcommand{\del}[1]{\old{#1}} 
	\newcommand{\ins}[1]{{\textcolor{blue}{\uline{#1}}}} 
	\newcommand{\ugh}[1]{{\textcolor{red}{\uwave{#1}}}} 
	\newcommand{\chg}[2]{{\textcolor{red}{\sout{#1}}}{\ra}\textcolor{blue}{\uline{#2}}} 
	 
	\newcommand{\fix}[1]{\bnote{FIX}{#1}}
}{
	\newcommand{\bnote}[2]{}
	\newcommand{\nb}[3]{}
	\newcommand{\old}[1]{}
	\newcommand{\del}[1]{}
	\newcommand{\ins}[1]{}
	\newcommand{\ugh}[1]{}
	\newcommand{\chg}[2]{}
	
	\newcommand{\fix}[1]{}
} 

\newcommand{\hide}[1]{}

\graphicspath{{}}

\newlist{RQ}{enumerate}{1}
\setlist[RQ]{label=RQ\arabic*:,leftmargin=3em}

\newlist{problems}{enumerate}{1}
\setlist[problems]{label=P\arabic*:,leftmargin=3em}

\newcommand{\commented}[1]{}

\newcommand{\etal}{\emph{et al.,}\xspace}
\newcommand{\ct}[1]{{\textsf{#1}}\xspace}

\usepackage{url}            
\makeatletter
\def\url@leostyle{%
  \@ifundefined{selectfont}{\def\UrlFont{\sf}}{\def\UrlFont{\small\sffamily}}}
\makeatother
\definecolor{main}{HTML}{828282}    
\definecolor{sub}{HTML}{E0E0E0}     

\tcbset{
    sharp corners,
    colback = white,
    before skip = 0.2cm,    
    after skip = 0.5cm      
}                           

\newtcolorbox{cbox}{
    enhanced, 
    boxrule = 0pt, 
    borderline = {0.75pt}{0pt}{main}, 
    borderline = {0.75pt}{2pt}{sub} 
}

\NewDocumentCommand{\bywhom}{m}{
  {\nobreak\hfill\penalty50\hskip1em\null\nobreak
   \hfill\mbox{\normalfont(#1)}%
   \parfillskip=0pt \finalhyphendemerits=0 \par}%
}

\NewDocumentEnvironment{pquotation}{m}
  {\begin{quoting}[
     indentfirst=true,
     leftmargin=\parindent,
     rightmargin=\parindent]\itshape}
  {\bywhom{#1}\end{quoting}}

\newcommand{\ph}{\textsc{Pharo}\xspace}
\newcommand{\compl}{\textsc{Complishon}\xspace}

\begin{document}
\copyrightyear{2026}
\copyrightclause{Copyright for this paper by its authors. Use permitted under Creative Commons License Attribution 4.0 International (CC BY 4.0).}

\conference{IWST 2026: International Workshop on Smalltalk Technologies, July 07--10, 2026, Plovdiv, Bulgaria}

\title{Alternative UX Extensions and Their Trade-offs for Code Completion in Pharo}

\author[1]{Mehdi Elkolei}[%
email=mehdi.elkolei.etu@univ-lille.fr,%
]

\author[1]{Omar AbedelKader}[%
email=omar.abedelkader@inria.fr,%
orcid=0009-0005-1339-5683,%
url=https://omarabedelkader.github.io%
]

\author[1]{Stéphane Ducasse}[%
orcid=0000-0001-6070-6599,
email=stephane.ducasse@inria.fr,
url=https://stephane.ducasse.free.fr/,
]

\address[1]{Univ. Lille, Inria, CNRS, Centrale Lille, UMR 9189 CRIStAL, Park Plaza, Parc scientifique de la Haute-Borne, 40 Av. Halley Bât A, 59650 Villeneuve-d’Ascq, France}

\begin{abstract}
Complishon is Pharo’s context-aware code completion engine, built on AST analysis, lazy candidate generation, and filter-based candidate selection. Its existing design already provides strong semantic completion, but several practical limits remain. Strict prefix matching is sensitive to small typing errors, framework prefixes often force redundant input, large completion menus are difficult to scan, and prefix-only matching does not support common camel-case abbreviations. This paper presents four extensions that address these limits: \emph{typo tolerance}, \emph{implicit prefix expansion}, \emph{grouped completion entries}, and \emph{camel-case matching}. For each extension, we describe the implementation, discuss alternative designs, and explain the trade-offs involved. The main contribution of the paper is to show that these improvements can be integrated into \compl while preserving its modular architecture and predictable behavior.
\end{abstract}


\maketitle

\section{Introduction}\label{introduction}
Code completion is useful only when it helps developers write code with less effort and less interruption. A completion system should not only return relevant candidates, but should also remain easy to understand, responsive, and stable during interaction \cite{IntelliJ}. This is especially important in live programming environments such as \ph, where completion operates on semantic information and must react quickly while the developer is editing code. \compl~\cite{Abed25a} is a heuristic, context-aware completion engine for \ph. It is built around AST analysis, lazy fetchers, and filter-based candidate construction. Rather than relying on a single monolithic algorithm, it uses the program context to guide completion and delegates candidate retrieval to specialized components. This design already provides a strong base for semantic completion. However, some practical limitations still appear in everyday use:

\begin{itemize}
\item First, strict prefix matching is unforgiving. A small typo can prevent the intended candidate from appearing at all. 
\item Second, framework and package naming conventions often require users to type prefixes that are already obvious from the surrounding context. 
\item Third, even when the appropriate candidates are present, long flat completion lists can be difficult to read and navigate, especially when many entries share similar names.
\item Fourth, prefix-only access does not support common abbreviation habits such as typing \ct{SC} for \ct{SortedCollection}.
\end{itemize}

This paper studies four extensions that address these problems: \emph{typo tolerance}, \emph{implicit prefix expansion}, \emph{grouped completion} entries, and \emph{camel-case} matching. These extensions do not all affect the same part of the system. Typo tolerance and camel-case matching extend the matching policy. Implicit prefix expansion changes the text used during matching while preserving the inserted identifier. Grouping changes the presentation of results without changing candidate retrieval. Together, they improve robustness, reduce unnecessary typing, and make completion results easier to browse. The goal of this paper is not only to describe these four features, but also to explain why they fit the architecture of \compl. The paper, therefore, treats each extension as a design choice: we present the implemented extension, compare it with realistic alternatives, and discuss its main strengths and limitations. In this way, the paper shows how a completion engine can be improved through targeted and modular changes rather than a complete redesign.

\section{Contribution}
The paper makes two contributions. First, it introduces four complementary extensions to \compl : typo-tolerant matching, implicit prefix expansion, grouped completion entries, and camel-case matching. These extensions address input errors, redundant framework prefixes, visually repetitive result lists, and abbreviation-based access to identifiers, respectively.
Second, the paper presents an architectural account of how these extensions can be integrated into an existing completion engine. The individual matching and presentation techniques are not themselves new. The contribution lies in their integration into \compl through localized changes to its filters, fetchers, entries, completion context, and user interface. This integration demonstrates that the behavior of a mature completion system can be extended without replacing its existing architecture or introducing a monolithic matching mechanism. For each extension, we describe its implementation, compare it with alternative designs, and analyze its benefits and limitations. The implementations are available as open-source projects to support replication, experimentation, and further development.

\section{Background}
\begin{figure}[!h]
\includegraphics[width=\linewidth]{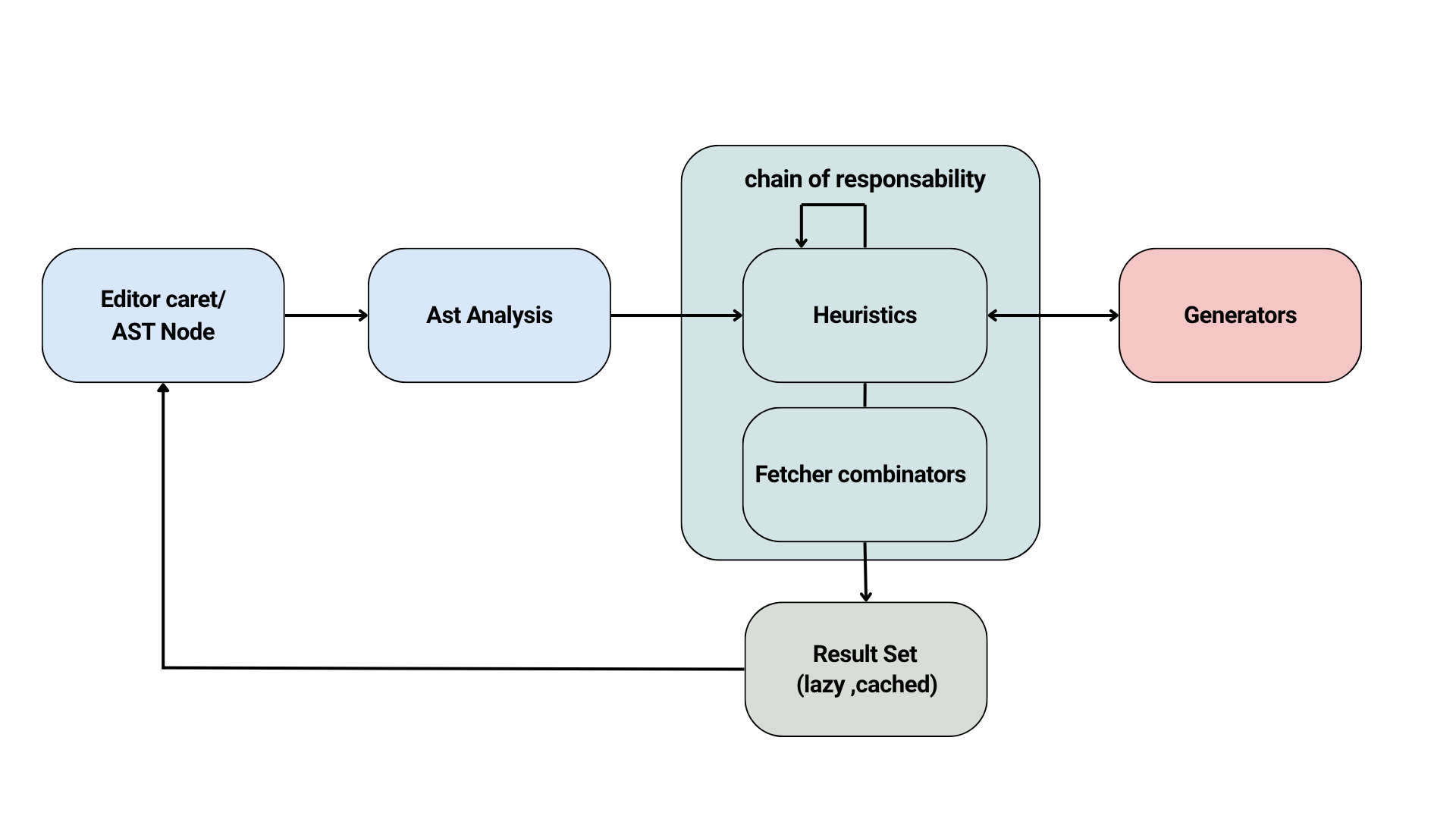}
\caption{\compl architecture overview \cite{Abed25a,icsme}.}\label{fig:architecture}
\end{figure}

\compl is \ph’s code completion engine. As illustrated in Figure~\ref{fig:architecture}, its completion pipeline is organized around three principal components: heuristics, fetchers, and result sets.

\paragraph{Heuristics.} Heuristics define a number of conditions based on the analysis of the Abstract Syntax Tree (AST) node corresponding to the current cursor location (editor caret). Once a heuristic is selected, it delegates the search for candidates to the fetchers. Heuristics are modular and specialized, each depending on a given context such as messages or variables, and are chained so as to focus first on self messages and instance variables before moving on to global variables and messages. They are tested one by one, using the chain of responsibility design pattern \cite{Gamm95a}, until the heuristic able to handle the current node is found.

\paragraph{Fetchers.} Once a heuristic has been selected, fetchers retrieve candidates that are relevant to the current program context. Candidate generation is lazy: fetchers use generators that produce entries incrementally and may stop once the requested number of candidates has been obtained. This design avoids constructing the complete candidate set when only a small result list is needed. Fetchers may also be combined or decorated to remove duplicate candidates, narrow the search space, or reuse existing candidate sources. Filters associated with the fetchers determine whether partially typed tokens match the generated candidates.

\paragraph{Result Set.} The result set stores the entries produced by the selected fetchers. It supports lazy retrieval and caching, allowing the completion interface to request additional candidates without restarting the entire completion process.

\section{Availability and Usage}\label{sec:availability}
The four extensions are implemented as separate open-source projects, whose repositories are referenced in the corresponding sections. This separation allows each mechanism to be studied and enabled independently. The extensions preserve \compl’s existing candidate-generation pipeline and introduce their behavior through localized subclasses, preferences, and user-interface components.

\section{Related Work}\label{related-work}

Research on code completion has shown that completion quality depends on more than candidate generation alone~\cite{Li21b}. Good completion also requires useful ranking, effective use of context, and low interaction cost for the developer. A first line of work studies how completion candidates should be ranked. Bruch \etal~\cite{Bruc09a} showed that ranking quality has a direct effect on how useful completion is in practice. Their work is important because it shifts attention from simply producing candidates to presenting the most relevant ones first. This idea is closely related to the present paper, especially for typo tolerance, where broader matching must still preserve understandable ordering. A second important direction models code as a structured and repetitive form of text. Hindle~\etal~\cite{Hind12a} argued that source code is natural in the sense that it contains regular and predictable patterns. This observation motivated many later systems that use statistical or language-model-based techniques for code suggestion. Hellendoorn~\etal~\cite{Hell19a} also showed that completion systems should be evaluated carefully in realistic settings, since strong reported results may not always translate into practical usefulness. These works are relevant here because they emphasize that completion should support real developer behavior rather than idealized input.

Another related area concerns identifier structure and naming conventions. Allamanis~\etal~\cite{Alla18a} reviewed machine learning approaches for source code and highlighted the importance of names and structural regularities in programming languages. This is directly relevant to camel-case matching, since camel-case identifiers already encode internal boundaries that can be used to support abbreviation-based completion. Context-aware completion is also strongly related to the present work. Several recent approaches~\cite{contextmodule2024,liao2023context} use wider contextual signals, repository structure, or surrounding code to improve suggestions. The same general idea supports implicit prefix expansion in \compl. The goal is not to guess arbitrary aliases, but to reduce redundant typing when the framework or package prefix is already strongly suggested by the local context. Recent work on package-aware completion in \ph also supports this direction by showing that repository and package information can improve completion for global names~\cite{Abed25a}. The usability of completion interfaces has also received increasing attention. Li~\etal~\cite{Li21b} argued that completion systems impose a hidden interaction cost, because developers must browse lists, inspect candidates, and decide what to accept. Wang\etal~\cite{Wang23b} likewise showed that developers care not only about accuracy, but also about the way completion tools behave in practice. These results are closely connected to the grouping extension proposed in this paper. Grouping does not change the underlying completion semantics, but it aims to reduce the cost of scanning large, visually repetitive result lists. Compared with this prior work, the contribution of the present paper is more modest and more local. It does not propose a new learned completion model or a new ranking framework. Instead, it studies four targeted extensions to an existing completion engine: typo tolerance, implicit prefix expansion, grouped presentation, and camel-case matching. The value of this work lies in showing how such improvements can be introduced within the modular architecture of \compl, while keeping the behavior understandable, configurable, and compatible with the existing system.

\section{Extension I: Typo Tolerance}

\subsection{Motivation and design alternatives}
Most completion systems start from prefix matching because it is simple, fast, and easy to understand. However, strict prefix matching is very sensitive to typing errors. If the user types \ct{Presnter} instead of \ct{Presenter}, or swaps two adjacent letters, the intended candidate is often removed from the result list even when the user input is very close to the target. This makes the interaction less smooth because the user must either fix the token first or continue without useful suggestions. This issue is well known in work on approximate string matching. Edit-distance models~\cite{damerau1964spelling,levenshtein1966binary} were introduced to capture common mistakes such as insertion, deletion, substitution, and transposition. Later a survey~\cite{navarro2001approximate} showed that these models are a practical basis for tolerant retrieval systems because they provide a clear way to measure similarity while remaining interpretable and configurable. In software tools, related work on code completion~\cite{Bruc09a,Hind12a,Hell19a} has also shown that better ranking and tolerance mechanisms can improve the usefulness of completion results in practice.

Given this, several alternatives can be considered.
\begin{itemize}
    \item \textbf{Alternative A: strict prefix only.} The simplest baseline is to keep the current prefix-only behavior.
    \begin{itemize}
        \item Advantages: It has the lowest computational cost, is easy to explain, and gives very stable rankings.
        \item Limitations: It is fragile under small typing mistakes and does not support recovery when the user input is close to the intended token but not an exact prefix.
    \end{itemize}

    \item \textbf{Alternative B: unrestricted fuzzy matching over all candidates.} A second option is to run approximate matching broadly and continuously over the full candidate set.
    \begin{itemize}
        \item Advantages: This can improve recall and make the system more tolerant to noisy inputs.
        \item Limitations: This approach is more expensive and more likely to return irrelevant items. In a programming setting, that risk is high because identifiers can be similar at the character level while being unrelated at the usage level. For keyword selectors, unrestricted fuzzy matching is especially risky because it can ignore structural distinctions that matter in the language.
    \end{itemize}

    \item \textbf{Alternative C: learned correction or usage-based reranking.} A third option is to use statistical or learned methods to predict the intended completion or reorder candidates from usage data. Prior work on code completion has shown that learning from examples or language regularities can improve ranking quality in many settings~\cite{Bruc09a,Hind12a}.
    \begin{itemize}
        \item Advantages: This may produce stronger rankings and adapt better to project-specific vocabulary or usage patterns.
        \item Limitations: It requires data collection, training, or a persistent state. It is also harder to explain and debug. For the present goal, it would introduce much more complexity than needed.
    \end{itemize}
\end{itemize}

\begin{figure}[!h]
\includegraphics[width=\linewidth]{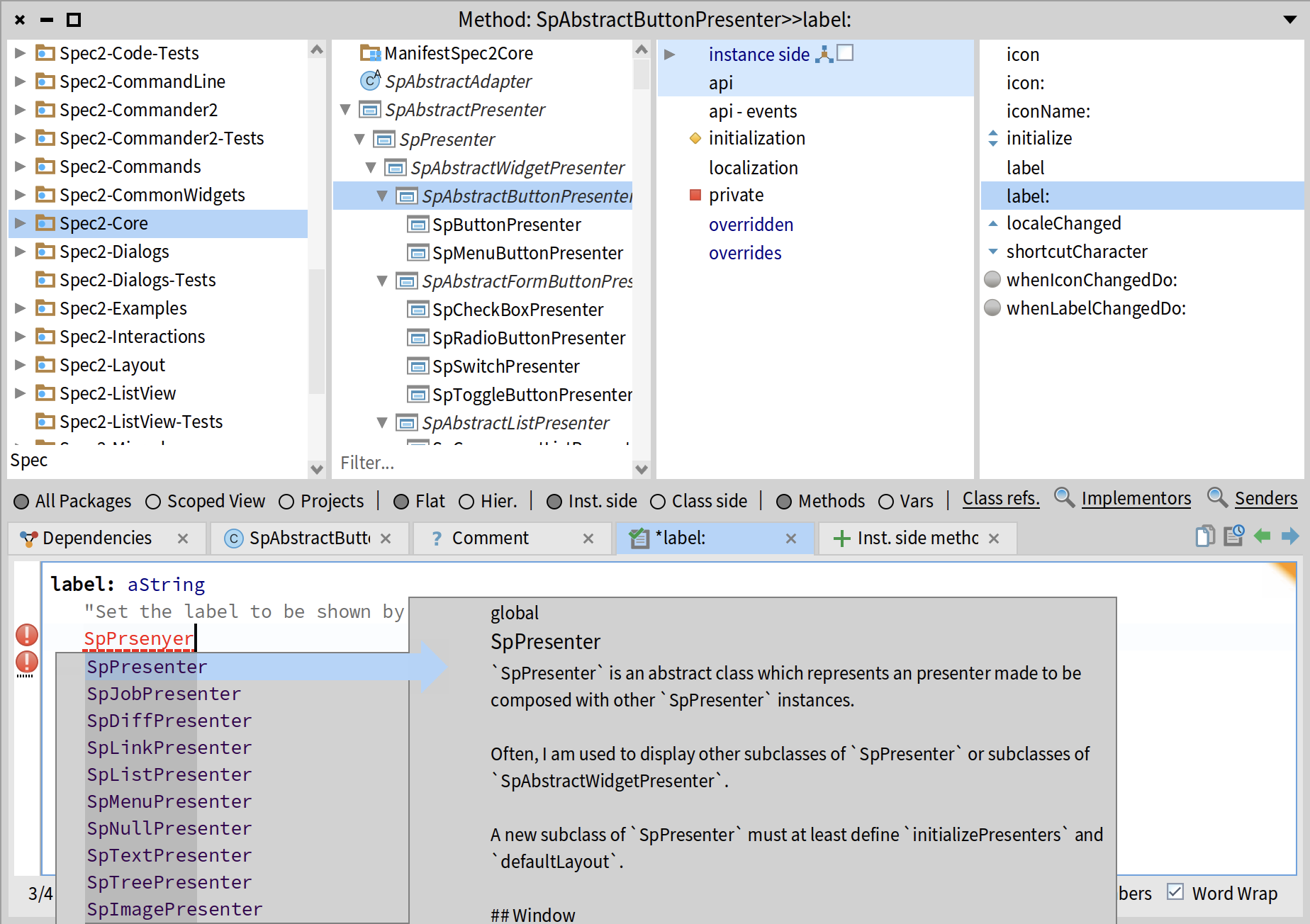}
\caption{Typo-tolerant completion retrieves the intended \texttt{SpPresenter}-related candidates despite an error in the typed token. }\label{fig:typo1}
\end{figure}

\subsection{Design goal}
The goal of this extension is not to replace prefix completion, but to extend it with limited typo tolerance (as shown in Figure~\ref{fig:typo1}). In other words, \compl should continue to behave like a normal prefix-based completion engine when the user types a correct prefix. Fuzzy matching~\cite{damerau1964spelling,levenshtein1966binary} should only help when strict prefix matching would otherwise fail or return too few useful results. This design choice keeps the interaction model stable. Users who type correct prefixes still obtain the expected results immediately, while users who make small typing mistakes can still recover without restarting the completion process.

\subsection{Implemented extension}
\begin{figure}[!h]
    \includegraphics[width=\linewidth]{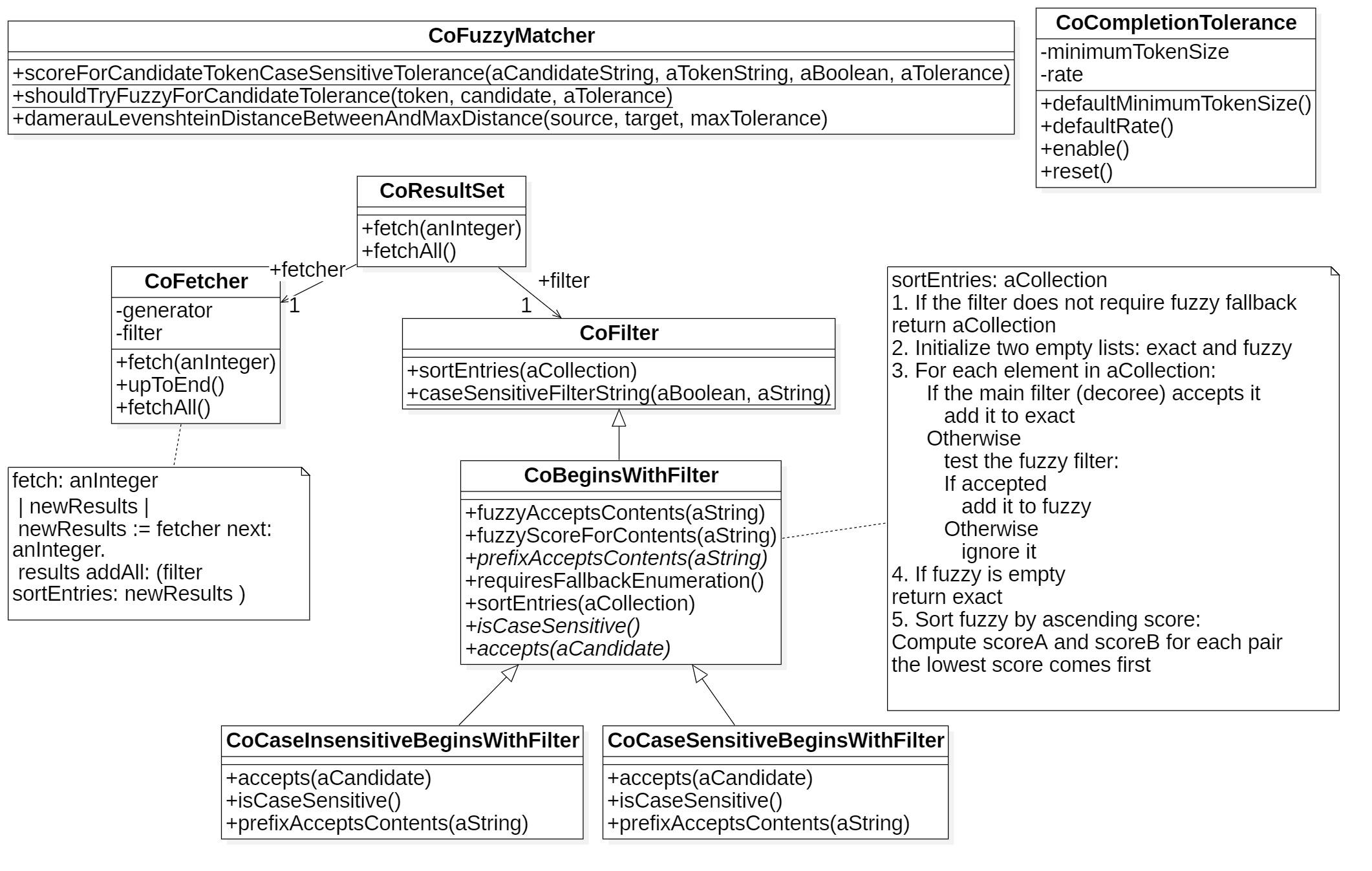}
    \caption{
Class diagram of the fuzzy matching fallback mechanism.}\label{fig:typo}
\end{figure}

The implementation introduces configurable typo tolerance through \ct{CoCompletionTolerance} and \ct{CoFuzzyMatcher} as shown in Figure~\ref{fig:typo}\footnote{\href{https://github.com/pharo-completion/typo-tolerance}{github.com/pharo-completion/typo-tolerance}}. The main idea is to preserve prefix matching as the primary acceptance rule and to use approximate matching only as a fallback when tolerance is enabled.

The extension has four main parts:

\paragraph{(1) Configurable tolerance.}
A global preference controls whether fuzzy completion is enabled and how strong it can be. The main parameters are a tolerance rate in \([0,1]\) and a minimum token size. This makes the feature adjustable: low tolerance values keep the behavior close to strict prefix completion, while higher values admit more candidates. A minimum token size is needed because very short inputs are too ambiguous and tend to generate noisy matches.

\paragraph{(2) Edit-distance matching.}
Fuzzy acceptance is based on Damerau--Levenshtein distance~\cite{damerau1964spelling,levenshtein1966binary}. This choice is suitable for interactive completion because it covers the most common typing mistakes:
\begin{itemize}
    \item insertion,
    \item deletion,
    \item substitution,
    \item transposition of adjacent characters.
\end{itemize}
Among classical string-similarity models, this distance is a good fit for completion because transpositions are common in real typing and should not block retrieval of an otherwise obvious target~\cite{navarro2001approximate}.

\paragraph{(3) Guardrails to reduce noise.}
A direct use of edit distance on all candidates would return too many weak matches. To avoid that problem, the implementation adds several guardrails:
\begin{itemize}
    \item fuzzy matching starts only after a minimum token length of two characters.
    \item selector structure is preserved by respecting keyword arity and colons;
    \item the first character must match;
    \item candidates are rejected early when the length difference is already larger than the allowed edit budget, computed as \(\lceil r \times \max(|token|, |candidate|) \rceil\), where \(r\) is the configured tolerance rate.
\end{itemize}

These constraints are important because \compl works on program identifiers and selectors, not on free text. In \ph, punctuation and keyword structure carry meaning, so a tolerant matcher must remain aware of that structure \cite{Zait20a}. The goal is therefore not maximum recall, but controlled recovery from small typing errors.

\paragraph{(4) Stable result ordering.}
Exact prefix matches remain first-class results. The completion menu separates exact prefix matches from fuzzy-only matches. Exact matches are shown first, while fuzzy matches are sorted by normalized similarity score. This policy keeps the ranking easy to understand: the system still rewards clean prefixes, and fuzzy results only appear as a secondary aid.

Filters now use fuzzy matching as a fallback once exact matching fails. The new mechanism, implemented in \ct{CoBeginsWithFilter} and driven by \ct{sortEntries}, will only work if the filter requires fallback enumeration : \ct{CoCompletionTolerance}'s rate is not zero and the token size is longer than the minimum token size.
For each candidate, \ct{CoFuzzyMatcher} computes a similarity score using the Damerau–Levenshtein distance between the token and the candidate; the algorithm determines whether the tokens are completely different by returning nil if the score is beyond the rate, otherwise the candidate is accepted as a fuzzy match. Once we have gathered enough candidates or processed all the information, \ct{complementResults} splits the results into two lists: exact matches accepted by the main filter, and fuzzy matches accepted only by the fuzzy fallback. The fuzzy list is then sorted by ascending score, and the two lists are combined, with exact matches placed before fuzzy ones.

\subsection{Discussion}
This extension fits \compl because it can be added mostly at the filter matching level. The fetchers just get configured with filters that accept a tolerance and edit distance. The fetchers do not need a new semantic model, a training phase, or usage history. They continue to produce candidates as before. The tolerance mechanism only decides whether a candidate should be accepted when prefix matching fails, and how fuzzy-only results should be ordered after exact ones. The change remains local, which reduces implementation risk and preserves the current architecture. It also makes the behavior easier to debug than a learned ranking system, since acceptance depends on explicit and inspectable rules.

Among these options, configurable edit-distance fallback is a good middle ground. It improves robustness to typing errors without changing the basic completion model of \compl. The approach remains lightweight, explicit, and compatible with the current system design. Its main weakness is that fuzzy matching adds some extra cost and may still admit a few marginal candidates. However, this risk is reduced by the use of \ct{minimumTokenSize}, first-character checking, selector-shape preservation, and exact-before-fuzzy ordering. In that sense, the extension is conservative by design: it adds tolerance, but only within clear limits. The expected benefit is better usability during interactive completion. Users should be able to recover from small mistakes without interrupting their flow, while still seeing the same exact-prefix behavior when they type a correct token. This follows the general direction of prior completion research, which argues that completion quality depends not only on candidate generation but also on how robustly the system supports realistic developer behavior~\cite{Bruc09a,Hell19a}.

\section{Extension II: Implicit Prefix Expansion}

\subsection{Motivation and Design Alternatives}
Many \ph libraries use naming conventions that encode package or framework identity directly in global names. In Spec2, for instance, many classes begin with the prefix \ct{Sp}. Such prefixes are useful because they improve namespace discipline and make framework membership visible in the source code. However, they also introduce redundancy during code completion. When the programming context already strongly suggests a given framework or package, requiring the developer to retype the full prefix is unnecessary repetition.

\begin{figure}[!h]
\includegraphics[width=\linewidth]{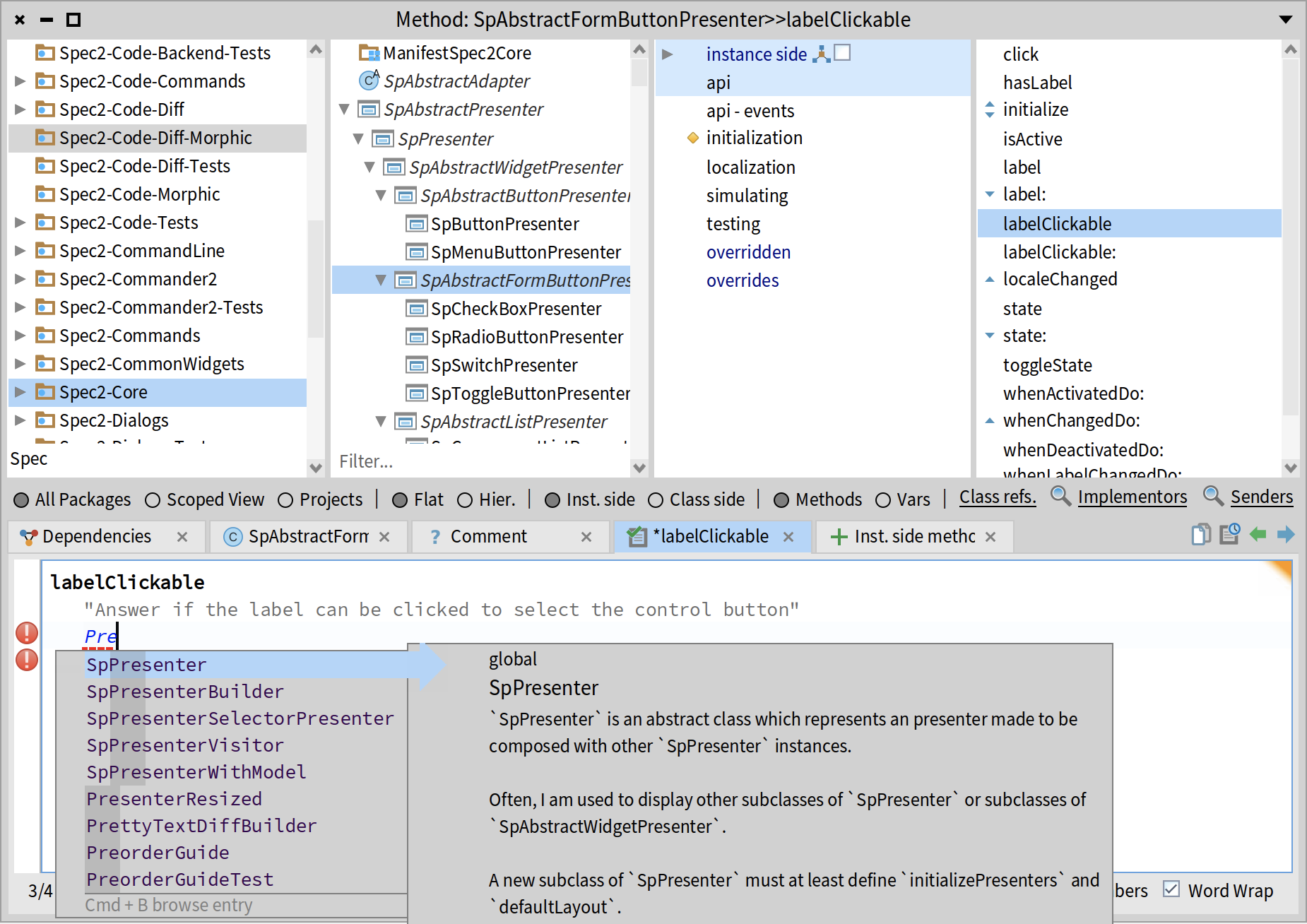}
\caption{ Implicit prefix expansion allows the suffix \texttt{Pre} to match \texttt{SpPresenter} when the \texttt{Sp} framework prefix is inferred from the current context.}\label{fig:prefix}
\end{figure}

The problem is that the completion engine may require users to repeat information that is already available in the surrounding context. Prior work on code completion~\cite{Bruc09a,Hind12a,contextmodule2024,liao2023context} has repeatedly shown that completion quality depends not only on lexical matching, but also on the ability to use contextual information effectively. In repository-aware and context-aware systems, broader environmental signals can improve both the relevance and the usability of suggestions~\cite{contextmodule2024,liao2023context}. A similar idea applies here at a smaller scale: if the local context already makes a framework prefix highly likely, then the completion process should be able to exploit that information (as shown in Figure~\ref{fig:prefix}).

Based on these considerations, several alternative solutions can be identified.
\begin{itemize}
    \item \textbf{Alternative A: no prefix inference.} The baseline is to require the user to type the explicit prefix that appears in the real global name.
    \begin{itemize}
        \item Advantages: This approach maximizes explicitness, introduces no hidden inference, and preserves purely lexical behavior.
        \item Limitations: It makes completion more repetitive in frameworks with systematic naming prefixes and does not exploit contextual information that is already available.
    \end{itemize}

    \item \textbf{Alternative B: explicit alias dictionaries.} Another option is to maintain explicit mappings such as \ct{Pre} $\rightarrow$ \ct{SpPresenter} or \ct{Txt} $\rightarrow$ \ct{SpTextPresenter}.
    \begin{itemize}
        \item Advantages: This can be powerful and customizable, and it may support project-specific abbreviations.
        \item Limitations: It requires maintenance, introduces the risk of inconsistency across users or teams, and scales poorly when conventions are structural rather than exceptional. It also moves the burden from the naming scheme to the maintenance of alias definitions.
    \end{itemize}

    \item \textbf{Alternative C: package-aware reordering without expanded matching.} A third possibility is to keep strict full-name matching but rank likely package-local entries higher in the result list. Recent \ph work on package-aware completion shows that repository structure can indeed improve ranking quality for global names~\cite{Abed25a}.
    \begin{itemize}
        \item Advantages: This approach is simpler than changing matching behavior and preserves explicit lexical acceptance.
        \item Limitations: It does not solve the central problem when the user omits the prefix entirely. It may improve ranking, but not acceptance. If \ct{Pre} does not lexically match \ct{SpPresenter}, reordering alone cannot recover the candidate.
    \end{itemize}
\end{itemize}

\subsection{Implemented extension}
\begin{figure}[!h]
    \includegraphics[width=\linewidth]{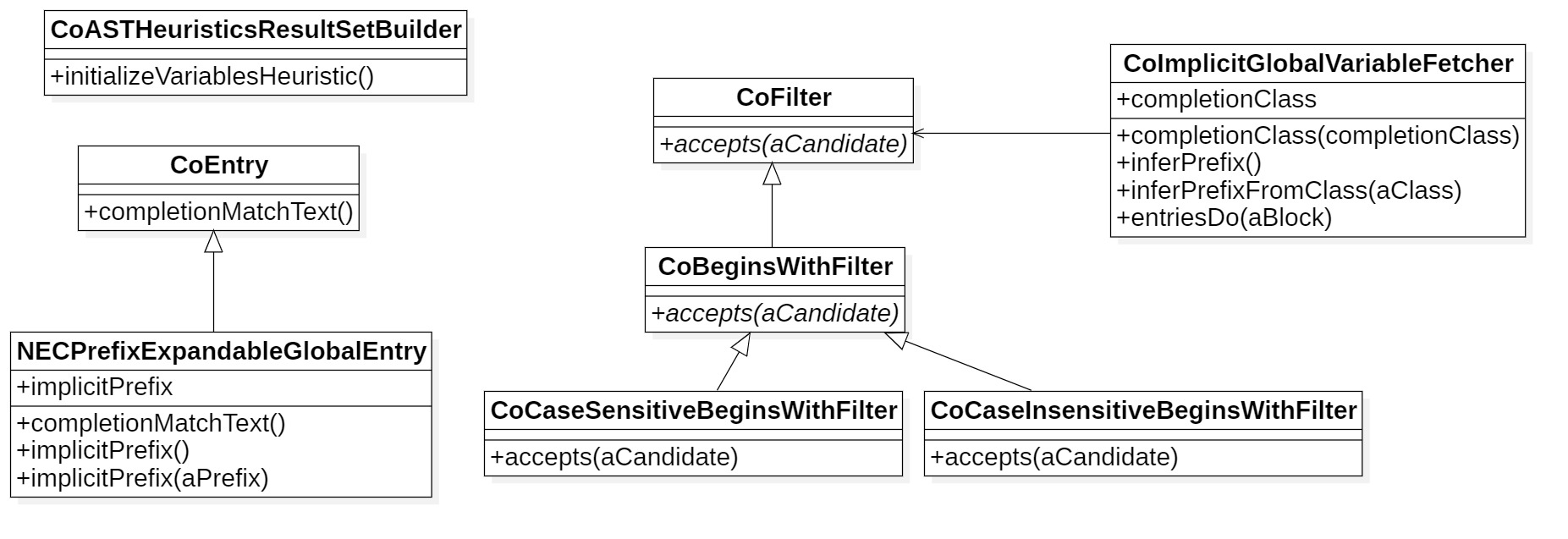}
    \caption{Class diagram of the implicit prefix expansion.}\label{fig:implicit}
\end{figure}

The implemented approach introduces \ct{NECPrefixExpandableGlobalEntry} as shown in Figure~\ref{fig:implicit}\footnote{\href{https://github.com/pharo-completion/implicit-prefix-matching}{github.com/pharo-completion/implicit-prefix-matching}}, a subclass of \ct{CoEntry}. The key idea is to distinguish between the inserted name of a completion candidate and the text used to match that candidate: instead of modifying the actual global identifier, the entry overrides \ct{completionMatchText} and exposes its inferred prefix through \ct{implicitPrefix}. To infer this prefix, a new fetcher, \ct{CoImplicitGlobalVariableFetcher}, uses information about the current context: through its \ct{completionClass} property, it examines the current package and all friendly packages to determine their most common prefix, computed by \ct{inferPrefixFromClass)} and exposed via \ct{inferPrefix)}. Once the prefix is found, it is concatenated with the user token, and the result is matched against other packages' classes as the fetcher iterates over candidates with \ct{entriesDo)}; matching itself is delegated to \ct{CoFilter} and its subclasses through \ct{accepts)}. When prefix expansion is enabled and an implicit prefix is known, the match text is computed as the suffix that follows the inferred prefix. Conceptually, this means that a user token such as \ct{Pre} can match \ct{SpPresenter} when the system infers \ct{Sp} from the current context. The inserted completion remains \ct{SpPresenter}; only the matching representation is shortened. This is an important design choice: the completion engine does not introduce aliases, duplicate global entries, or rewrite source-level identifiers. Instead, it changes only the representation used for acceptance during completion, making the mechanism lightweight and local while preserving the integrity of the inserted symbol.

\subsection{Design rationale}
The extension is based on three design principles.

\paragraph{(1) Separation between matching and insertion.}
The system should distinguish the text that is matched from the text that is ultimately inserted into the source code. This separation allows the completion engine to be flexible at selection time while remaining exact at insertion time. From an engineering perspective, this is safer than introducing a parallel aliasing mechanism because the real identifier is never altered.

\paragraph{(2) Context-sensitive reduction of redundant typing.}
The purpose of implicit prefix expansion is to reduce repeated input when the package or framework identity is already strongly implied. This is consistent with the broader goal of context-aware completion systems, which aim to reduce developer effort by exploiting information already present in the working environment~\cite{Bruc09a,contextmodule2024,liao2023context}.

\paragraph{(3) Preservation of source-level explicitness.}
Although the matcher may ignore an inferred prefix during acceptance, the system still inserts the original global name. As a result, source code remains explicit, stable, and consistent with library naming conventions. The extension, therefore, improves completion ergonomics without weakening the naming discipline of the codebase.

\subsection{Discussion}
This extension fits \compl particularly well because completion entries are already modeled as objects with their own behavior. Extending an entry so that it can expose a specialized matching representation is therefore more local than introducing a global alias table or a more invasive rewriting subsystem. This object-level design has two practical advantages. First, it keeps the implementation modular: the change is attached to entry behavior rather than spread across the whole completion pipeline. Second, it remains reversible and easy to reason about, because the inserted representation is unchanged. In this sense, implicit prefix expansion follows the architecture of \compl by enriching candidate behavior rather than replacing the overall completion model. Implicit prefix expansion is a strong usability feature when naming conventions are regular, and the relevant package context is clear. Its main strength is that it reduces redundant typing without changing the actual identifier that is inserted into the program.

Its main weakness is the possibility of ambiguity. If multiple frameworks expose classes with similar suffixes, removing the inferred prefix from the matching representation may broaden the candidate space more than intended. For that reason, the feature should remain optional and context-sensitive. A preference flag is therefore appropriate because it allows the mechanism to be enabled only when the surrounding environment provides a sufficiently reliable package signal.

\subsection{Expected impact}
The expected result is a smoother completion workflow in prefix-heavy frameworks. Developers should be able to request classes such as \ct{SpPresenter} by typing only the meaningful suffix when the framework prefix is already implied by the context. This aligns with the general direction of modern completion research, where contextual information is used to reduce developer effort and improve the practical usefulness of completion systems~\cite{Bruc09a,contextmodule2024,liao2023context,Abed25a}.

\section{Extension III: Grouping Entries}

\subsection{Motivation and Design Alternatives}

Even when candidate matching is correct, completion lists may still be difficult to scan efficiently. In frameworks with strong naming regularity, completion often produces long runs of visually similar entries. A flat list is simple and familiar, but it does not always scale well from the perspective of human attention. When many candidates share the same leading characters, users must repeatedly inspect nearly identical prefixes before reaching the suffixes that actually distinguish one entry from another.

The problem is therefore not primarily algorithmic. The completion engine may already be retrieving relevant entries. The difficulty lies in presentation and navigation. Recent work~\cite{Li21b,Wang23b} on code completion has pointed out that the usefulness of completion systems depends not only on prediction accuracy, but also on the cost of browsing, interpreting, and accepting suggestions. In particular, Li~\etal~\cite{Li21b} argue that completion systems have a hidden interaction cost that comes from list browsing and candidate selection, while practitioner reports confirm that usability and display quality are central concerns in practice~\cite{Li21b,Wang23b}.

Accordingly, several alternative solutions can be identified.
\begin{itemize}
    \item \textbf{Alternative A: flat ranked list.} The default approach is to retain the existing design and display all relevant completions in one ranked list, without any extra grouping or visual structure.
    \begin{itemize}
        \item Advantages: This design is familiar, easy to implement, and supports straightforward keyboard navigation.
        \item Limitations: It can become visually overloaded on large result sets. Repeated textual prefixes consume screen space and increase scanning effort, especially when many results are lexically similar. This is precisely the kind of hidden browsing cost emphasized in recent completion research~\cite{Li21b}.
    \end{itemize}

    \item \textbf{Alternative B: semantic grouping by category or fetcher.} Another option is to group entries according to semantic categories such as globals, locals, selectors, classes, or package origin.
    \begin{itemize}
        \item Advantages: This can provide meaningful structure when textual similarity is weak, and it may help users reason about the provenance of suggestions.
        \item Limitations: It requires a more intrusive interface redesign and depends on stable metadata being exposed by the completion pipeline. In addition, semantic categories do not always align with the immediate visual problem the user faces, namely distinguishing highly similar names in a dense list.
    \end{itemize}

    \item \textbf{Alternative C: coarse alphabetical buckets.} A simpler design would partition entries into broad buckets such as \ct{A--C}, \ct{D--F}, or by first letter.
    \begin{itemize}
        \item Advantages: This is easy to implement and predictable.
        \item Limitations: It is usually too coarse to help with realistic completion tasks. Such buckets are not sensitive to the token the user has typed and therefore do not directly address the local redundancy that arises in prefix-heavy frameworks.
    \end{itemize}
\end{itemize}

\begin{figure}[!h]
\includegraphics[width=\linewidth]{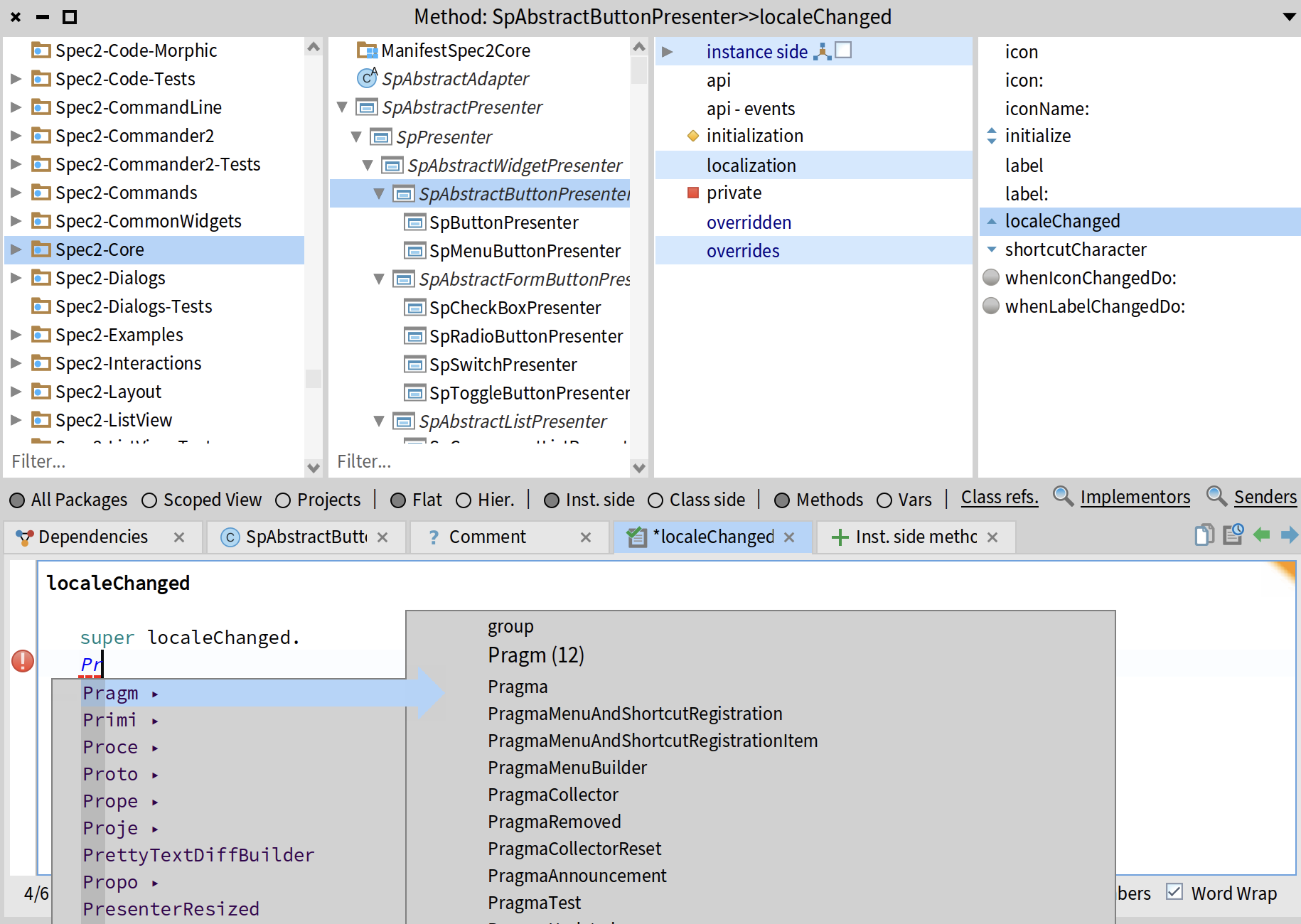}
\caption{Prefix-based grouping replaces a long flat candidate list with navigable groups of entries sharing a textual prefix.}\label{fig:grouping}
\end{figure}

\subsection{Implemented extension}
\begin{figure}[!h]
    \includegraphics[width=\linewidth]{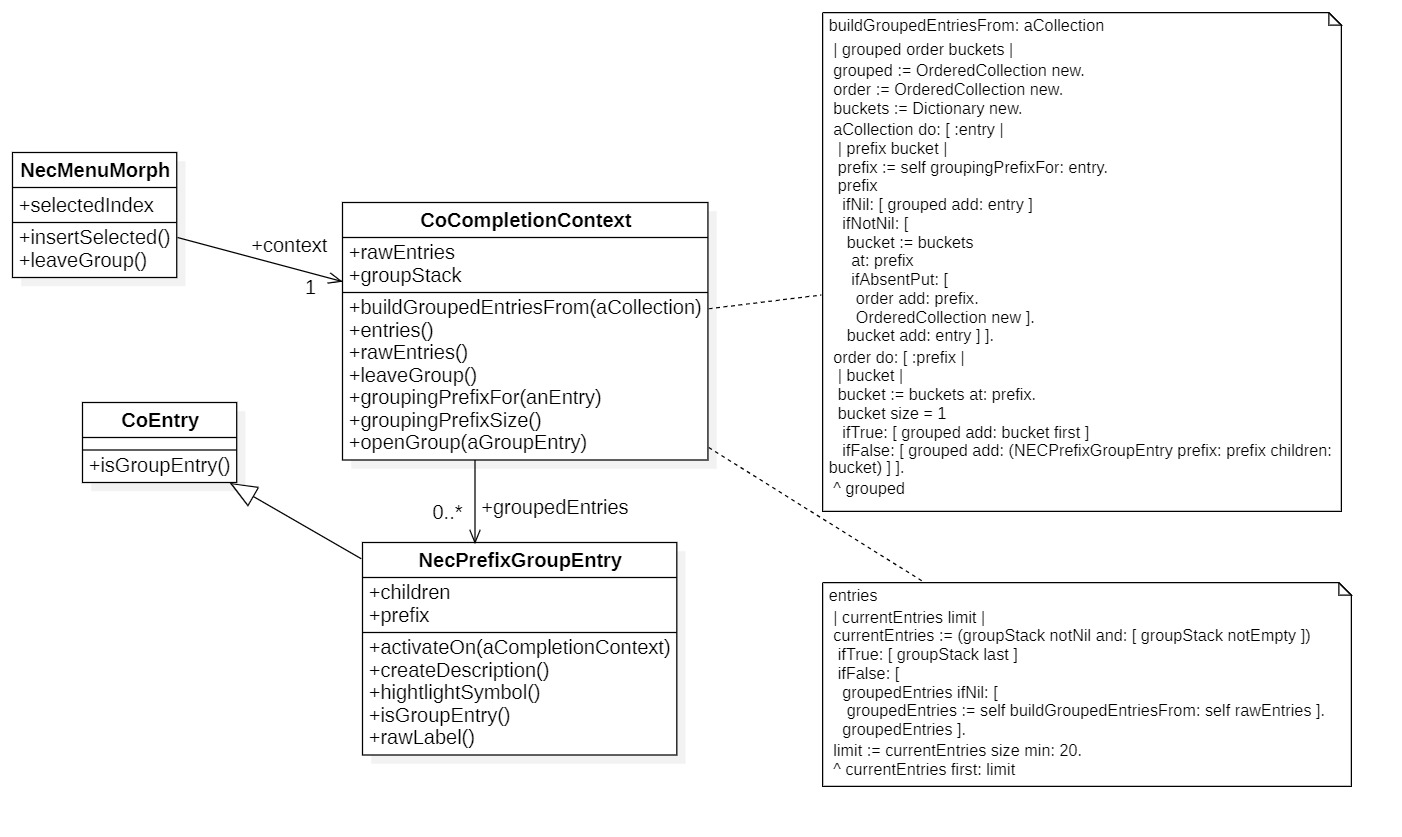}
    \caption{Class diagram of the grouped presentation mechanism.}\label{fig:groupedClass}
\end{figure}

The proposed extension introduces grouped presentation as shown in Figure~\ref{fig:grouping} through \ct{CoCompletionContext} and \ct{NecPrefixGroupedEntry} as shown in Figure~\ref{fig:groupedClass}\footnote{\href{https://github.com/pharo-completion/group-entries}{github.com/pharo-completion/group-entries}}. After computing the candidates, they are stored in \ct{CoCompletionContext} as \ct{rawEntries}. When grouping is enabled, these raw completion entries are partitioned according to a shared textual prefix using \ct{buildGroupedEntriesFrom:}, which will group elements starting with the same prefix into the same group, determined by \ct{groupingPrefixFor:)}. The grouping prefix size is configurable via \ct{groupingPrefixSize()}, and the effective prefix length is defined as the maximum of the configured grouping size and the size of the token already typed by the user; there is no grouping for candidates that have a length less than the minimum allowed. Entries that do not belong to a true cluster remain plain completion entries, while only multi-entry clusters are converted into group entries, each represented as a \ct{NecPrefixGroupEntry} with its own \ct{prefix} and \ct{children}.
A group entry is not a synthetic completion result. It is a navigational node that represents a subset of already available entries, exposed through \ct{isGroupEntry()} and rendered via \ct{createDescription()} and \ct{rawLabel()}. Activating such a node, through \ct{activateOn(aCompletionContext)}, opens a sublist containing its children by pushing onto \ct{CoCompletionContext}'s \ct{groupStack} and calling \ct{openGroup(aGroupEntry)}. \ct{NecMenuMorph} handles the interaction with the user, meaning showing the groups and navigating between items and groups: the completion menu is extended accordingly so that users can enter groups via \ct{insertSelected()} and leave them via \ct{leaveGroup()}. In this way, grouping is implemented as a lightweight hierarchical interface layered on top of the existing completion mechanism.

\subsection{Design rationale}
The extension follows three main design principles.

\paragraph{(1) Presentation should be separated from retrieval.}
Grouping does not attempt to discover new candidates or alter the semantics of matching. Instead, it reorganizes already-fetched entries into a form that is easier to scan. This distinction is important because the problem being addressed is one of presentation rather than search.

\paragraph{(2) Grouping should reflect the structure already visible to the user.}
Since \compl is fundamentally prefix-oriented, grouping by shared leading text is a natural extension of the existing interaction model. The grouped structure remains closely aligned with the way users already interpret completion results, especially in frameworks where naming conventions are systematic.

\paragraph{(3) Additional structure should remain lightweight.}
The goal is not to transform the completion menu into a full semantic browser. Instead, the design introduces only one extra navigation layer. This keeps the mechanism simple enough for interactive use while still reducing visual clutter in large result sets.

\subsection{Discussion}

This extension fits \compl because it is implemented at the context and menu levels rather than inside fetchers. This is an appropriate design choice because grouping is not about finding additional entries; it is about presenting existing entries more effectively. By distinguishing \ct{rawEntries} from \ct{groupedEntries}, \compl preserves the integrity of the original result set while enabling a second layer of organization. This separation also keeps the architecture modular. Fetchers continue to produce completion candidates exactly as before. Grouping is applied only after retrieval, which reduces implementation risk and makes the feature easier to disable or adapt. In architectural terms, the extension extends the user interface without changing the underlying completion semantics. Prefix-based grouping is a practical compromise between a completely flat list and a much heavier semantic browser. Its main strengths are simplicity, locality, and consistency with the prefix-oriented behavior already present in \compl. By reducing long runs of visually similar entries into navigable clusters, it lowers scanning effort without changing the underlying result set.

Its main limitation is that textual grouping is only a partial proxy for conceptual grouping. Two entries may be semantically related without sharing a useful prefix, and two entries may share a prefix while representing unrelated concepts. For that reason, the extension should not be understood as a general classification system. Rather, it is a targeted interface improvement for completion scenarios where naming regularity already dominates the visible structure of the candidate list.

\subsection{Expected impact}
The expected benefit is improved readability of dense completion menus and lower effort during candidate selection. In practice, this should be most useful in frameworks with systematic naming conventions, where large families of entries differ mainly in their suffixes. More broadly, the extension follows the direction suggested by recent completion research and practitioner feedback: a useful completion system is not defined only by what it can retrieve, but also by how effectively users can inspect and act on its results~\cite{Li21b,Wang23b,Abed25a}.

\section{Extension IV: Camel-Case Matching}

\subsection{Motivation and Design Alternatives}
Developers often rely on abbreviations when interacting with code completion systems. In object-oriented environments with long and descriptive identifiers, it is common to type compact forms such as \ct{SC} for \ct{SortedCollection}, \ct{MNU} for \ct{MessageNotUnderstood}, or \ct{SpP} for \ct{SpPresenter}. These abbreviations exploit the internal structure of identifiers, particularly the boundaries indicated by camel-case capitalization. Prefix-based completion does not support this behavior well because it assumes that the user input corresponds to a literal prefix of the candidate. As a result, many valid and natural abbreviations are rejected, even though they clearly reflect the structure of the intended identifier. Prior work on code completion~\cite{Hind12a,Alla18a} and identifier modeling has shown that developers rely heavily on naming conventions and structural regularities when writing and searching for code. Supporting these patterns can therefore improve both usability and discoverability.

Given this, several alternative approaches can be considered.
\begin{itemize}
    \item \textbf{Alternative A: prefix-only completion.} The baseline approach is to accept only literal prefixes.
    \begin{itemize}
        \item Advantages: It is fast, deterministic, and requires no additional scanning.
        \item Limitations: It does not support common abbreviation patterns and reduces usability for long identifiers.
    \end{itemize}

    \item \textbf{Alternative B: acronym-only matching.} A simpler approach is to match only the first letter of each camel-case segment.
    \begin{itemize}
        \item Advantages: This is easy to implement and efficient, and it works well for highly regular names.
        \item Limitations: It is less flexible than prefix-based segment matching and cannot handle partial abbreviations such as \ct{SortC} for \ct{SortedCollection}.
    \end{itemize}

    \item \textbf{Alternative C: general subsequence matching.} Another option is to accept any ordered subsequence of characters.
    \begin{itemize}
        \item Advantages: This provides very high recall and can match a wide range of abbreviations.
        \item Limitations: It is often too permissive, leading to noisy results. Because it ignores structural boundaries, it becomes harder to interpret and rank matches effectively.
    \end{itemize}

    \item \textbf{Alternative D: learned abbreviation expansion.} A more advanced approach is to use statistical or machine learning models to infer likely expansions from user input.
    \begin{itemize}
        \item Advantages: Such models can capture domain-specific abbreviations and usage patterns.
        \item Limitations: They require training data, introduce additional system complexity, and reduce transparency compared to rule-based approaches~\cite{Alla18a}.
    \end{itemize}
\end{itemize}

\begin{figure}[!h]
\includegraphics[width=\linewidth]{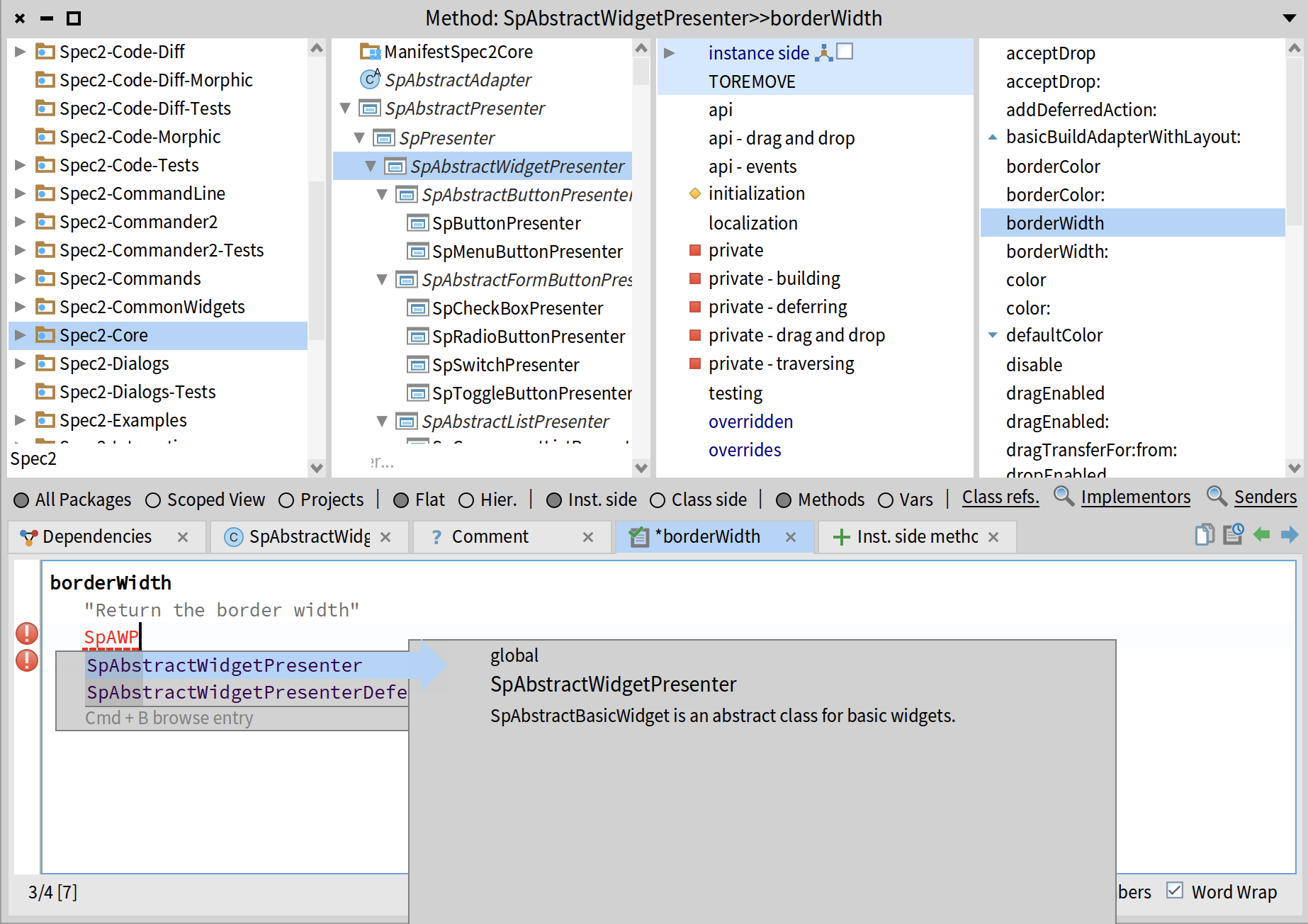}
\caption{Camel-case matching retrieves \texttt{SpAbstractWidgetPresenter} from an abbreviation aligned with its capitalization boundaries.}\label{fig:camel}
\end{figure}

\subsection{Implemented extension}
The proposed extension introduces optional camel-case matching for global and class completions as shown in Figure~\ref{fig:camel}. The approach extends the notion of prefix matching so that it can operate over the internal structure of identifiers rather than only over their literal beginning.\footnote{\href{https://github.com/pharo-completion/camel-case-matching}{github.com/pharo-completion/camel-case-matching}}

The implementation relies on two main ideas.

\paragraph{(1) Entry-level matching capabilities.}
Completion filters can query candidates to determine whether they support specialized matching strategies such as camel-case matching. This avoids concentrating all matching logic in a single global filter and instead distributes responsibility to the entries themselves. This design keeps the system modular and allows different kinds of entries to expose different matching behaviors.

\begin{figure}[!h]
    \includegraphics[width=\linewidth]{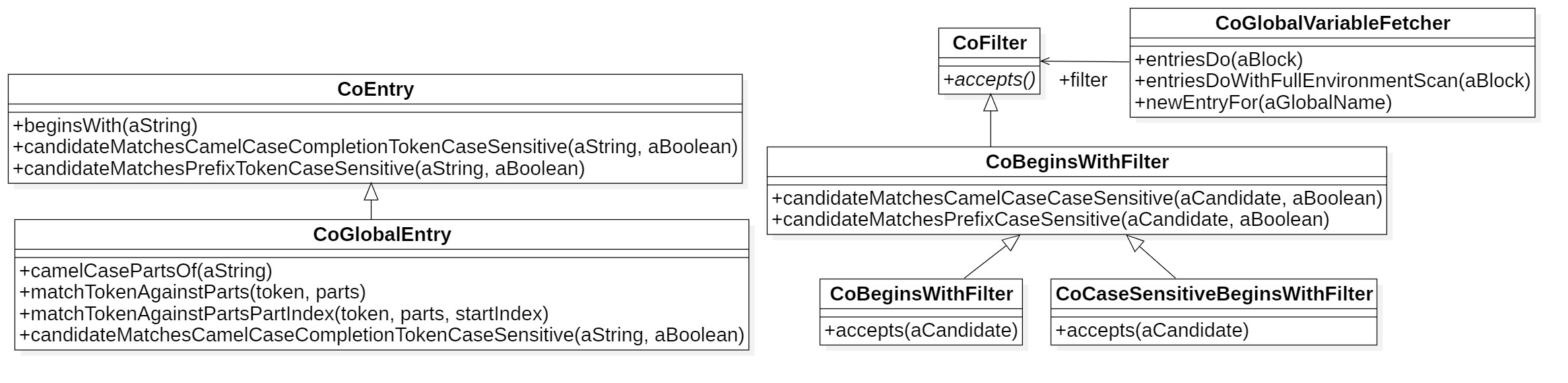}
    \caption{Class diagram of the camel-case matching mechanism.}\label{fig:camelClass}
\end{figure}

\paragraph{(2) Structured camel-case decomposition.}
\ct{CoGlobalEntry} implements camel-case matching by decomposing an identifier into segments based on capitalization boundaries, via \ct{camelCasePartsOf:)}. The matching process then checks whether the user token can be consumed as a sequence of prefixes across these segments using recursive matching, implemented as \ct{matchToken:againstParts:} and \ct{matchToken:againstParts:partIndex:}, and exposed to filters through \ct{candidate:matchesCamelCaseCompletionTokenCaseSensitive:}; after decomposing the candidate identifier into segments, the token is matched against these segments by consuming successive prefixes: at each segment, we try progressively longer prefixes of that segment and check whether the token begins with that prefix. If it matches, we consume that portion of the token and continue with the next segment; if no prefix of the current segment leads to a match, we backtrack and try starting the match from the next segment instead. The token is considered matched only once it has been fully consumed this way. This mechanism is used as a fallback at the filter level, as shown in Figure~\ref{fig:camelClass}: it is only evaluated when the strict prefix match fails. For example, the token \ct{SC} matches \ct{SortedCollection} because it corresponds to the initial letters of the segments \ct{Sorted} and \ct{Collection}. Similarly, partial matches such as \ct{SortC} are accepted because they align with segment prefixes. This extension is intentionally stricter than arbitrary subsequence matching. It does not accept any ordered set of characters. Instead, it requires that the abbreviation follows the structural boundaries already encoded in the identifier. This constraint preserves a clear relationship between the user input and the candidate, making matches easier to understand and rank. When camel-case matching is enabled, candidate retrieval may require a broader scan of the environment: \ct{CoGlobalVariableFetcher} falls back from its default \ct{entriesDo)} to \ct{entriesDoWithFullEnvironmentScan)}. Prefix-based enumeration is no longer sufficient because valid matches may occur at positions beyond the initial characters. This additional cost represents the main trade-off of the approach.

\subsection{Design rationale}
The extension is guided by three principles.

\paragraph{(1) Exploiting identifier structure.}
Camel-case identifiers encode semantic boundaries through capitalization. Leveraging these boundaries allows the completion engine to interpret abbreviations that are already natural to developers. This aligns with observations that naming conventions play a central role in code comprehension and navigation~\cite{Alla18a}.

\paragraph{(2) Controlled flexibility.}
The approach extends prefix matching without fully relaxing it. By requiring alignment with camel-case segments, the system remains more constrained than general subsequence matching. This helps maintain precision and reduces the risk of irrelevant matches.

\paragraph{(3) Modular extensibility.}
By delegating matching behavior to completion entries, the system remains extensible. New matching strategies can be introduced without modifying a single central filter, which is consistent with the architecture of \compl.

\subsection{Discussion}
Camel-case matching fits well with the overall philosophy of \compl. The system remains symbolic and structure-based rather than statistical. It continues to operate on identifier form and explicit rules, without requiring training data or probabilistic models. At the same time, the extension broadens the interpretation of user input in a way that reflects real developer behavior. Instead of assuming that the token is a literal prefix, the system allows it to represent a structured abbreviation. This extension improves expressiveness while preserving the transparency of the matching process. Camel-case matching is a strong extension for expert users because it transforms identifier structure into a compact access mechanism. It allows developers to express intent concisely while still aligning with the naming conventions of the codebase.

The main trade-off is performance. Since matching is no longer restricted to literal prefixes, the system may need to consider a larger set of candidates during filtering. However, this cost remains moderate compared to more permissive approaches such as general subsequence matching, and it preserves a clear structural interpretation of user input.

\subsection{Expected impact}
The expected benefit is improved efficiency during code navigation and completion, especially for experienced users who rely on abbreviations. By allowing compact representations of structured identifiers, the system reduces typing effort while preserving meaningful alignment between user input and candidate names. This follows the broader direction of completion research, which emphasizes leveraging the regularities of code to improve developer productivity~\cite{Hind12a,Alla18a}.

\section{Discussion}
Although the four extensions address different limitations of \compl, they are not independent. Their main contribution appears not only in isolation, but also in the way they interact at the levels of matching, ranking, and presentation. Prior work on code completion~\cite{Bruc09a,Hind12a,Li21b,Wang23b} has shown that completion quality depends not only on candidate generation, but also on ranking quality, contextual relevance, and the overall interaction cost imposed on the developer. The four extensions proposed here should therefore be understood as complementary improvements to the completion experience rather than as isolated features. 

\paragraph{Tolerance, recall, and ranking.}
Typo tolerance improves recall by recovering candidates that would be lost under strict prefix matching. However, this benefit comes with a known trade-off: broader acceptance can also introduce borderline or weakly relevant candidates. For this reason, ranking becomes especially important once fuzzy matching is enabled. The decision to preserve exact prefix matches ahead of fuzzy-only matches is therefore central to the design, since it maintains predictability while still allowing recovery from typing errors. This is consistent with work showing that ranking quality strongly affects the practical usefulness of completion systems and that browsing poor suggestions carries a real interaction cost for developers~\cite{Bruc09a,Li21b}. 

\paragraph{Inference and non-literal matching.}
Implicit prefix expansion and camel-case matching both increase the expressive power of completion, but they do so through different mechanisms. Implicit prefix expansion relies on contextual inference: it allows the engine to omit a redundant framework prefix when that prefix is already strongly suggested by the surrounding environment. Camel-case matching, in contrast, relies on the internal structure of identifiers and interprets abbreviations through capitalization boundaries. In both cases, accepted user input is no longer restricted to the literal beginning of the final identifier. This makes completion more flexible and closer to common developer practice, while also increasing the need for understandable ordering and stable defaults. Such a balance between flexibility and predictability is in line with both repository-aware completion research and practitioner expectations for completion tools~\cite{Bruc09a,Wang23b}.

\paragraph{Architectural separation of concerns.}
A central architectural result of this work is that these extensions do not belong to a single implementation layer. Typo tolerance is mainly a filtering and ranking concern. Implicit prefix expansion is primarily an entry-level concern because it changes how a candidate should be matched without changing the inserted identifier. Grouping is a presentation concern, located at the context and menu levels. Camel-case matching spans both entries and filters because it requires structural matching behavior together with compatible selection logic. This separation is an important strength of \compl. It suggests that completion systems benefit from modular extension points rather than monolithic matching logic, which is consistent with the broader software-engineering value of local and explainable design decisions in developer tools~\cite{Bruc09a,Wang23b}. A recurrent issue common to all extensions emerges during the presentation of completion candidates: when the user hovers over a suggestion, the corresponding documentation is displayed. In many cases, this documentation is extensive and obscures substantial portions of the underlying source code, which can significantly impair usability, especially during refactoring or maintenance activities. This limitation primarily affects the overall user experience (UX) and lies outside the intended scope of our extension, as it exists within \compl itself. Therefore, it should be regarded as an additional potential area for future improvement.

These four extensions suggest that completion quality should not be understood only as a question of finding candidates. A useful completion system must also decide how tolerant matching should be, how contextual structure should be exploited, how abbreviations should be interpreted, and how result sets should be presented to the user. In that sense, the contribution of this work is both functional and architectural: it improves the interaction experience while also showing that \compl provides suitable extension points for future experimentation. This view is compatible with prior research emphasizing that code completion should be evaluated not only by raw prediction ability, but also by its effect on developer effort and usability~\cite{Li21b,Wang23b}. Table~\ref{tab:extension-summary} summarizes these four extensions by comparing their implemented approaches, main advantages, and main limitations.

\section{Summary Table}
\begin{longtable}{p{0.12\linewidth}p{0.25\linewidth}p{0.26\linewidth}p{0.26\linewidth}}
\caption{Summary of the four extensions}\label{tab:extension-summary}\\
\toprule
\textbf{Extension} & \textbf{Implemented approach} & \textbf{Main advantages} & \textbf{Main limitations} \\
\midrule
\endfirsthead

\toprule
\textbf{Extension} & \textbf{Implemented approach} & \textbf{Main advantages} & \textbf{Main limitations} \\
\midrule
\endhead

Typo tolerance &
Damerau--Levenshtein fallback with configurable tolerance and structural guardrails &
Improves robustness to typing errors; preserves exact-prefix priority; integrates naturally with existing filtering behavior &
Introduces additional computation; may admit marginal matches if tolerance is configured too aggressively \\ \hline

Implicit prefix expansion &
Entry-level suffix matching after contextual prefix inference &
Reduces redundant typing in prefix-heavy frameworks; preserves the original inserted identifier; remains lightweight and local to entry behavior &
May introduce ambiguity when several frameworks expose similar suffixes; depends on sufficiently reliable contextual inference \\ \hline

Grouping entries &
Prefix-based hierarchical grouping at the context and menu levels &
Reduces visual clutter; improves browsing in naming-regular frameworks; does not change candidate retrieval &
Textual grouping is only an approximation of semantic grouping; introduces an additional navigation step \\ \hline

Camel-case matching &
Structural abbreviation matching based on camel-case segmentation &
Supports common expert abbreviations; respects identifier structure; improves access to long names &
May require broader environment scans; more expensive than literal prefix matching \\

\bottomrule

\end{longtable}

\section{Conclusion}
This work introduced four complementary extensions to \compl; Typo tolerance, implicit prefix expansion, grouping , camel-case matching. These extensions address different limitations of the completion experience, ranging from robustness to input errors to browsing cost and support for  common abbreviation habits.

Beyond these individual benefits, the results also highlight an important architectural point: \compl can be improved substantially without abandoning its original design. The system does not need to be replaced by a fundamentally different completion paradigm in order to become more expressive and more helpful. Instead, targeted changes at the appropriate architectural layers are sufficient to improve both robustness and usability. This conclusion aligns with earlier completion research showing that meaningful gains can come from improving ranking, contextualization, and interaction design rather than only replacing the entire completion mechanism~\cite{Bruc09a,Li21b}

The implemented extensions are not necessarily the most aggressive or most powerful solutions that could be imagined. More permissive fuzzy matching, richer semantic grouping, or learned ranking models might offer stronger behavior in some situations. However, the approaches proposed here provide a better balance for a live programming environment such as \ph. They remain configurable, understandable, and well aligned with the current decomposition of \compl into entries, filters, contexts, and fetchers. This emphasis on explainability and usability is also consistent with practitioner feedback on completion tools, which values not only power but also predictability and smooth interaction~\cite{Wang23b}

\bibliography{rmod,others,us}

@article{Alla18a,
  author = {Allamanis, Miltiadis and Barr, Earl T and Devanbu, Premkumar and Sutton, Charles},
  title = {A survey of machine learning for big code and naturalness},
  journal = {ACM Computing Surveys (CSUR)},
  volume = {51},
  pages = {81},
  publisher = {ACM},
  year = {2018},
  number = {4}}

@inproceedings{Bruc09a,
  author = {Bruch, Marcel and Monperrus, Martin and Mezini, Mira},
  title = {Learning from examples to improve code completion systems},
  booktitle = {Proceedings of the 7th joint meeting of the European software engineering conference and the ACM SIGSOFT symposium on the foundations of software engineering},
  pages = {213--222},
  year = {2009}}

@misc{ContextS,
  title = {{ContextS}},
  url = {http://www.swa.hpi.uni-potsdam.de/downloads/index.html},
  key = {ContextS},
  note = {http://www.swa.hpi.uni-potsdam.de/downloads/index.html}}

@book{Gamm95a,
  author = {Erich Gamma and Richard Helm and Ralph Johnson and John Vlissides},
  title = {Design Patterns: Elements of Reusable Object-Oriented Software},
  publisher = {Addison-Wesley},
  year = {1995}}

@inproceedings{Hell19a,
  author = {Hellendoorn, Vincent J and Proksch, Sebastian and Gall, Harald C and Bacchelli, Alberto},
  title = {When code completion fails: A case study on real-world completions},
  booktitle = {2019 IEEE/ACM 41st International Conference on Software Engineering ({ICSE})},
  pages = {960--970},
  year = {2019},
  organization = {IEEE}}

@inproceedings{Hind12a,
  author = {Hindle, Abram and Barr, Earl T and Su, Zhendong and Gabel, Mark and Devanbu, Premkumar},
  title = {On the naturalness of software},
  booktitle = {Software Engineering ({ICSE}), 2012 34th International Conference on},
  pages = {837--847},
  year = {2012},
  organization = {IEEE}}

@inproceedings{Li21b,
  author = {Li, Jingxuan and Huang, Rui and Li, Wei and Yao, Kai and Tan, Weiguo},
  title = {{Toward Less Hidden Cost of Code Completion with Acceptance and Ranking Models}},
  booktitle = {International Conference on Software Maintenance and Evolution (ICSME)},
  year = {2021}}

@inproceedings{Wang23b,
  author = {Wang, Chaozheng and Hu, Junhao and Gao, Cuiyun and Jin, Yu and Xie, Tao and Huang, Hailiang and Lei, Zhenyu and Deng, Yuetang},
  title = {How Practitioners Expect Code Completion?},
  booktitle = {European Software Engineering Conference and Symposium on the Foundations of Software Engineering (ESEC/FSE)},
  year = {2023}}

@inproceedings{Abed25a,
  author = {AbedelKader, Omar and Ducasse, St{\'e}phane and Zaitsev, Oleksandr and Robbes, Romain and Polito, Guillermo},
  title = {Package-Aware Approach for Repository-Level Code Completion in Pharo},
  booktitle = {IWST 2025: International Workshop on Smalltalk Technologies},
  publisher = {CEUR Workshop},
  year = {2025},
  url = {https://hal.science/hal-05446902},
  doi = {10.48550/ARXIV.2601.05617},
  hal-id = {hal-05446902}
}

@techreport{Zait20a,
  author = {Zaitsev, Oleksandr and Ducasse, St{\'e}phane and Anquetil, Nicolas},
  title = {Characterizing Pharo Code: A Technical Report},
  year = {2020},
  type = {Technical Report},
  institution = {Inria Lille Nord Europe - Laboratoire CRIStAL - Universit\'e de Lille ; Arolla},
  tagnicolasa = {other},
  month = jan,
  hal-id = {hal-02440055},
  hal-url = {https://hal.inria.fr/hal-02440055/document},
  tagnicolasa = {other}}

@article{damerau1964spelling,
author = {Damerau, Fred J.},
title = {A technique for computer detection and correction of spelling errors},
year = {1964},
journal = {Communications of the ACM},
}

@article{levenshtein1966binary,
  author    = {Vladimir I. Levenshtein},
  title     = {Binary Codes Capable of Correcting Deletions, Insertions, and Reversals},
  journal   = {Soviet Physics Doklady},
  year      = {1966}
}

@article{navarro2001approximate,
  author    = {Gonzalo Navarro},
  title     = {A Guided Tour to Approximate String Matching},
  journal   = {ACM Computing Surveys},
  year      = {2001},
}

@article{contextmodule2024,
  author={Guan, Zhanming and Liu, Junlin and Liu, Jierui and Peng, Chao and Liu, Dexin and Sun, Ningyuan and Jiang, Bo and Li, Wenchao and Liu, Jie and Zhu, Hang},
  title     = {ContextModule: Improving Code Completion via Repository-level Contextual Information},
  journal   = {arXiv},
  year      = {2024}
}

@article{liao2023context,
  author    = {Dianshu Liao and Shidong Pan and Qing Huang and Xiaoxue Ren and Zhenchang Xing and Huan Jin and Qinying Li},
  title     = {CodGen: A Repository-Level Code Generation Framework for Code Reuse With Local-Aware, Global-Aware, and Third-Party-Library-Aware},
  journal   = {Transactions on Software Engineering},
  year      = {2024}
}

@misc{IntelliJ,
  author = {{JetBrains}},
  title = {Code completion in IntelliJ IDEA},
  year = {2026},
  url = {https://www.jetbrains.com/help/idea/auto-completing-code.html},
  note = {Accessed 2026-05-06}
}

@inproceedings{icsme,
  author = {Kier, Kilian and Giagnorio, Alessandro and Abedelkader, Omar and Zaitsev, Oleksandr and Peharz, Robert and Robbes, Romain and Bavota, Gabriele and Ducasse, Stéphane},
  title = {Teaching LLMs a Low-Resource Language: Enhancing Code Completion in Pharo},
  booktitle = {International Conference on Software Maintenance and Evolution (ICSME)},
  doi = {10.48550/arXiv.2607.04939},
  year = {2026}}

\end{document}